\documentclass[acmsmall,screen,nonacm]{acmart}

\setcopyright{none}

\acmConference{}{}{}
\acmYear{}
\acmDOI{}
\acmISBN{}
\acmPrice{}
\acmSubmissionID{}
\renewcommand{\footnotetextcopyrightpermission}[1]{}

\usepackage{xspace}
\usepackage{booktabs}
\usepackage{graphicx}
\usepackage{cleveref}
\usepackage{array}
\usepackage[table]{xcolor}
\usepackage{threeparttable}

\begin{document}

\newcommand{\oursystem}{\textsc{TreeMind}\xspace}
\newcommand{\eg}{\textit{e.g.}\xspace}
\newcommand{\ie}{\textit{i.e.}\xspace}
\newcommand{\red}{\textcolor{red}{...}\xspace}

\title{\oursystem: Automatically Reproducing Android Bug Reports via LLM-empowered Monte Carlo Tree Search}

\author{Zhengyu Chen}
\email{zychen@stu.ahu.edu.cn}
\affiliation{
\institution{Anhui University}
\city{Hefei}
\state{Anhui}
\country{China}
}

\author{Zhaoyi Meng}
\email{zymeng@ahu.edu.cn}
\affiliation{
\institution{Anhui University}
\city{Hefei}
\state{Anhui}
\country{China}
}
\authornote{Corresponding author}

\author{Wenxiang Zhao}
\email{zhaowx98@mail.ustc.edu.cn}
\affiliation{
\institution{University of Science and Technology of China}
\city{Hefei}
\state{Anhui}
\country{China}
}

\author{Wansen Wang}
\email{23762@ahu.edu.cn}
\affiliation{
\institution{Anhui University}
\city{Hefei}
\state{Anhui}
\country{China}
}

\author{Wenchao Huang}
\email{huangwc@ustc.edu.com}
\affiliation{
\institution{University of Science and Technology of China}
\city{Hefei}
\state{Anhui}
\country{China}
}

\author{Jie Cui}
\email{cuijie@mail.ustc.edu.cn}
\affiliation{
\institution{Anhui University}
\city{Hefei}
\state{Anhui}
\country{China}
}

\author{Hong Zhong}
\email{zhongh@ahu.edu.cn}
\affiliation{
\institution{Anhui University}
\city{Hefei}
\state{Anhui}
\country{China}
}

\author{Yan Xiong}
\email{yxiong@ustc.edu.com}
\affiliation{
\institution{University of Science and Technology of China}
\city{Hefei}
\state{Anhui}
\country{China}
}


\begin{abstract}
Automatically reproducing Android app crashes from textual bug reports is challenging, particularly when the reports are incomplete and the modern UI exhibits high combinatorial complexity.
Existing approaches based solely on reinforcement learning or large language models (LLMs) exhibit limitations in such scenarios.
They struggle to infer unobserved steps and reconstruct the underlying user action sequences to navigate the vast UI interaction space, primarily due to limited goal-directed reasoning and planning.
We present \oursystem, a novel technique that integrates LLMs with an adapted Monte Carlo Tree Search (MCTS) algorithm to achieve strategic UI exploration in bug reproduction.
To the best of our knowledge, this is the first work to combine external decision-making with LLM semantic reasoning for reliable and accurate reproduction processes.
We formulate the reproduction task as a target-driven search problem, leveraging MCTS as the core planning mechanism to iteratively refine action sequences.
To enhance MCTS with semantic reasoning, we introduce two LLM-guided agents with distinct roles: Expander generates top-\textit{k} promising actions based on the current UI state and exploration history, while Simulator estimates the likelihood that each candidate action leads toward successful reproduction by additionally leveraging dynamic environment feedback.
By incorporating multi-modal UI inputs and tailored prompting strategies, \oursystem performs feedback-aware navigation that identifies essential user actions and incrementally reconstructs reproduction paths.
We evaluate \oursystem on a dataset of 93 real-world Android bug reports from three widely-used benchmarks. 
Experimental results show that it significantly outperforms four state-of-the-art baselines, including ReBL, ReActDroid, AdbGPT, and ReproBot, in reproduction success rate.
Ablation studies further demonstrate the effectiveness of each proposed strategy.
The evaluations indicate that integrating LLM reasoning with MCTS-based planning is a compelling direction for automated bug reproduction.
\end{abstract}

\maketitle


\section{Introduction}
Mobile apps become an integral part of nearly every aspect of modern life.
As reported by Statista, over 2 million Android apps are currently available in the Google Play Store~\cite{statista2025apps}.
In this competitive landscape, it is increasingly important for developers to ensure app quality through effective maintenance and timely bug fixing.
A widely cited survey indicates that 88\% of app users would abandon an app if they encountered bugs or glitches~\cite{Bolton2017AppAbandonment}.
As a result, developers are expected to identify and resolve these functionality issues promptly to retain users.
To support this, bug reports serve as a primary source of information to understand observed failures~\cite{feng2024prompting}.

To fix a reported crash, developers must first reproduce it based on the crash-triggering procedures (\eg, the sequence of user interactions) in the corresponding report.
However, a major challenge developers face is that the lack of critical details in bug reports substantially  increases the complexity of bug reproduction~\cite{huang2025one}.
Not all submitters, \ie, users who report crashes, follow the issue-reporting instructions or provide detailed procedures for reproducing the crashes~\cite{johnson2022empirical}.
Without detailed reproduction steps, developers must spend considerable time manually diagnosing the issues.

Many approaches have been proposed to automatically reproduce bug reports~\cite{zhang2023automatically,zhao2019recdroid,zhao2022recdroid+,fazzini2018automatically}.
Specifically, they extract entities from steps to reproduce (\ie, S2Rs) and then match them with the app UI to replay the reported bug.
However, it is non-trivial to accurately and completely extract S2Rs from bug reports using natural language processing techniques.
Meanwhile, explicitly matching bug reports with the app's UI elements may fail to capture reproduction steps that are omitted from the reports.
Moreover, the approaches based on dynamic exploration~\cite{zhao2019recdroid} or human-defined heuristics~\cite{fazzini2018automatically} may suffer from high overhead and limited performance in practice.

Recent years have witnessed the emergence of various bug report reproduction methods based on Large Language Models (LLMs)~\cite{feng2024prompting,wang2024feedback,huang2025one}.
These methods rely on advanced capabilities of LLMs to understand bug reports, analyze UI information of apps, and then drive the app exploration.
Although these methods outperform earlier approaches in terms of effectiveness and efficiency, a critical problem remains unresolved: The combinatorial complexity of modern UIs and the incompleteness of bug reports make LLM-only approaches inadequate for identifying viable reproduction steps.
Specifically, the growing complexity of current apps leads to a combinatorial explosion in the number of possible UI interaction sequences.
Meanwhile, the incompleteness of bug reports often necessitates the exploration of a much broader set of potential interaction paths.
Without explicit guidance or external feedback, LLMs have limited capability in systematically exploring and validating UI transitions, making them ill-suited for navigating the vast and uncertain interaction space.
While recent work~\cite{wang2024feedback} incorporates feedback to improve LLM decision-making, such feedback provides limited support for global planning and tends to be locally reactive, which makes the exploration process prone to compounding errors and resulting in less directed navigation.

Monte Carlo Tree Search (MCTS) has demonstrated remarkable reasoning and search capabilities in various domains, \eg, game playing~\cite{silver2017mastering} and automated planning~\cite{chekroun2024mbappe}.
It is an iterative tree-based algorithm that effectively balances exploration and exploitation when searching large and complex state spaces.
This is accomplished through a four-stage process: selection, expansion, simulation, and backpropagation, which incrementally  refine the search toward satisfactory solutions.
In our context, reproducing Android bugs from natural language reports is a challenging search problem in a vast and partially observable UI interaction space.
It requires identifying a sequence of user actions that can trigger the bug.
However, due to the complexity of the space, such sequences are often difficult to discover.
Given MCTS's strengths in navigating such environment, it is well-suited for guiding this search process.

We propose and implement \oursystem, a novel technique for automatic reproduction of Android bug reports via LLM-empowered MCTS.
Our primary goal is to achieve accurate bug reproduction even in the presence of incomplete bug reports, under a reasonable time constraint.
To this end, \oursystem employs an adapted MCTS algorithm integrated with an LLM-based multi-agent framework, enabling effective UI exploration.
MCTS serves as the core planning mechanism for the search aimed at triggering bugs.
During each MCTS iteration, \oursystem progressively interprets crash context and infer appropriate UI actions through LLM-guided agents enhanced with advanced prompt engineering techniques.
To further support more reliable reproduction decisions, \oursystem leverages both textual (\eg, structured UI descriptions) and visual (\eg, UI screenshots) inputs as complementary modalities for advanced LLMs.





There are two technical challenges in applying MCTS for our work: (i) When a bug report lacks sufficient detail, the space of possible actions in certain UI states can become extremely large, especially when the actions involve free-form text inputs or gestures with continuous parameters.
This makes it infeasible for MCTS to blindly expand the search tree from such states, often resulting in inefficient UI exploration.
(ii) Monte Carlo rollouts suffer from both the vast action space and the interaction latency imposed by real-time UI feedback, which significantly slow down the simulation process.
Consequently, most rollouts fail to reach terminal states where the target bug is triggered within a limited time budget, leading to suboptimal guidance during UI exploration.


To address the above challenges, we design two LLM-guided agents with distinct roles, \ie, Expander and Simulator, which are respectively aligned with the expansion and simulation stages in MCTS.
For the first challenge, to avoid exploring the intractably large action space, the Expander agent expands the top-\textit{k} candidate actions that are most likely to reproduce the bug report according to the current UI state and the exploration history.
This strategy promotes more targeted exploration by constraining the branching factor of the Monte Carlo search tree, striking a practical balance between efficiency and effectiveness.
For the second challenge, to overcome the difficulty of obtaining reward signals through full rollouts, the Simulator agent directly estimates the likelihood of reproducing the bug report, based on the context of UI exploration and each one-step look-ahead rollout result from the expanded actions.
This likelihood is transformed into a proxy reward, which is backpropagated to guide future search decisions.

To evaluate the effectiveness of \oursystem, we perform comprehensive experiments on a real-world dataset comprising 93 real-world Android bug reports from three widely-used benchmarks.
Compared to four state-of-the-art bug reproduction tools (\ie, ReActDroid~\cite{huang2025one}, ReBL~\cite{wang2024feedback}, AdbGPT~\cite{feng2024prompting}, and ReproBot~\cite{zhang2023automatically}), \oursystem achieves the highest success rate of 64.52\% success rate, outperforming ReActDroid (45.16\%), ReBL (40.86\%), AdbGPT (34.41\%), and ReproBot (31.18\%).
Ablation studies further validate the contribution of each proposed strategy.
The evaluations indicate that integrating LLM reasoning with MCTS-based planning is a compelling direction for automated bug reproduction.
Our main contributions are summarized as follows:
\begin{enumerate}
    \item We present \oursystem, the first work, to the best of our knowledge, that 
    combines external decision-making with LLM semantic reasoning for goal-directed bug reproduction.
    \item Experimental results demonstrate that \oursystem significantly outperforms four state-of-the-art Android bug report reproduction tools (\ie, ReActDroid, ReBL, AdbGPT, and ReproBot) in reproduction success rate.
    \item  We will make the implementation and dataset of \oursystem publicly available upon acceptance to facilitate future research work.
\end{enumerate}

\begin{figure}[t]
    \centering
    \includegraphics[width=\textwidth]{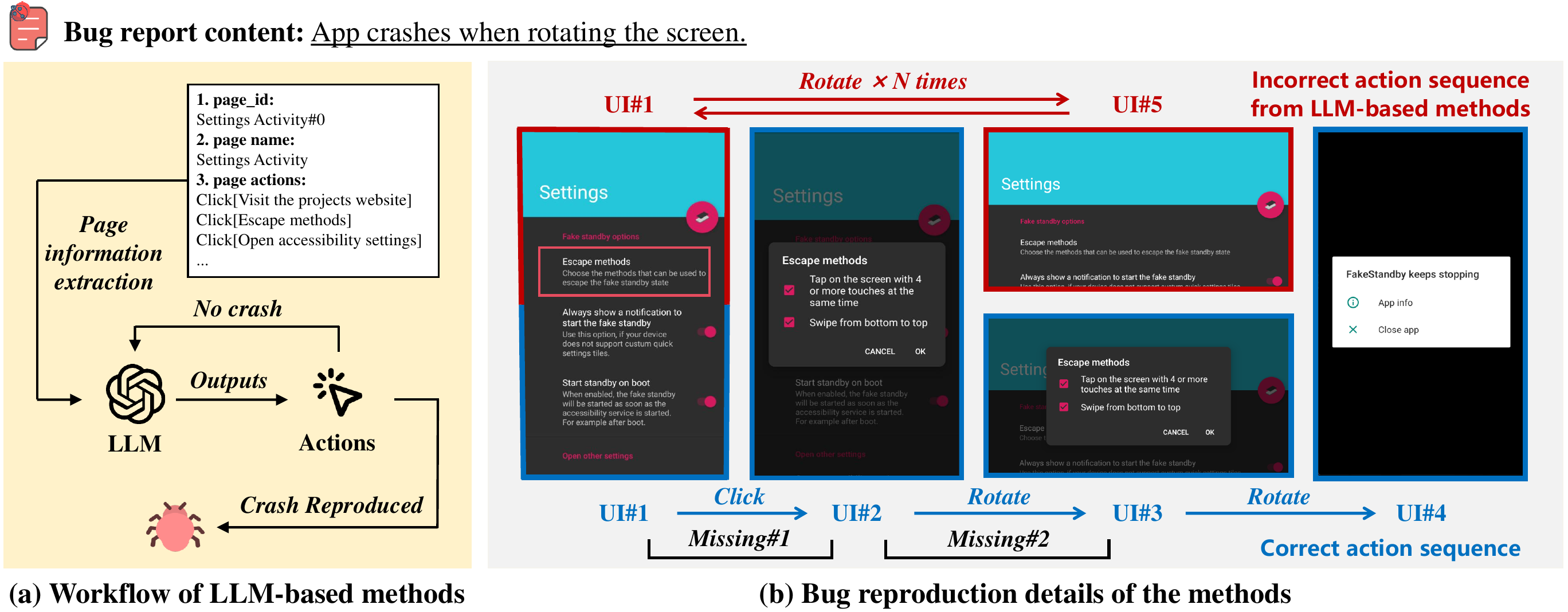}
    \caption{Motivating example}
    \label{fig:case}
\end{figure}

\section{Motivation}\label{sec:motivation}
Reproducing crashes in Android apps poses a persistent challenge, mainly due to incomplete bug reports. 
To enhance the effectiveness of bug reproduction, many researchers have explored diverse automated methods.
For example, ReproBot~\cite{zhang2023automatically} uses Q-learning to automate UI exploration, while AdbGPT~\cite{feng2024prompting}, ReBL \cite{wang2024feedback}, and ReActDroid~\cite{huang2025one} leverage LLMs to generate reproduction actions from bug reports.
\Cref{fig:case}(a) illustrates the overall workflow of the LLM-based methods.


When bug reports omit essential execution steps, existing methods have difficulty generating correct actions for reproducing target bugs.
As exemplified at the top of \Cref{fig:case}, the bug report consists of only a single sentence instructing users to rotate the screen to trigger the crash of an app named \textit{FakeStandby}.
However, as shown in \Cref{fig:case}(b), this description omits two critical interaction actions: (i) Clicking the \textit{"escape methods"} option on the screen (\ie, \textit{Missing\#1}), and (ii) rotating the screen while keeping the dialog open (\ie, \textit{Missing\#2}).
These missing actions have only a weak semantic connection to the bug report, making them difficult to infer by LLMs.
In this situation, ReproBot tends to repeat actions such as rotation in the early stages and gradually broadens its search as negative feedback accumulates.
AdbGPT, ReBL and ReActDroid try to rotate the screen repeatedly, thereby causing continuous transitions between UI\#1 and UI\#5 along the red arrows.
Although ReActDroid introduces a novel crash localization mechanism, it still struggles to identify the crashing page under such conditions.
As a result, in our experiments, all these methods fail to reproduce the crash within the given time constraints, \ie, 30 minutes.

This limitation underscores the need for a more adaptive exploration mechanism that can infer the missing user actions not explicitly mentioned in the bug report.
Rather than gradually broadening the search within the app or relying solely on LLM-generated suggestions, an effective solution should systematically explore the interaction space while progressively exploiting more promising interaction paths.
We propose to integrate LLMs with MCTS.
In this design, LLMs are used to reason about plausible missing actions and evaluate their outcomes based on the context of the bug reproduction process, while MCTS provides a structured decision-making process that discovers diverse action sequences and refines the search toward the most likely reproduction path.
This synergy enables the recovery of essential actions, allowing reproduction from incomplete reports.

\begin{figure}[t]
    \centering
    \includegraphics[width=\textwidth]{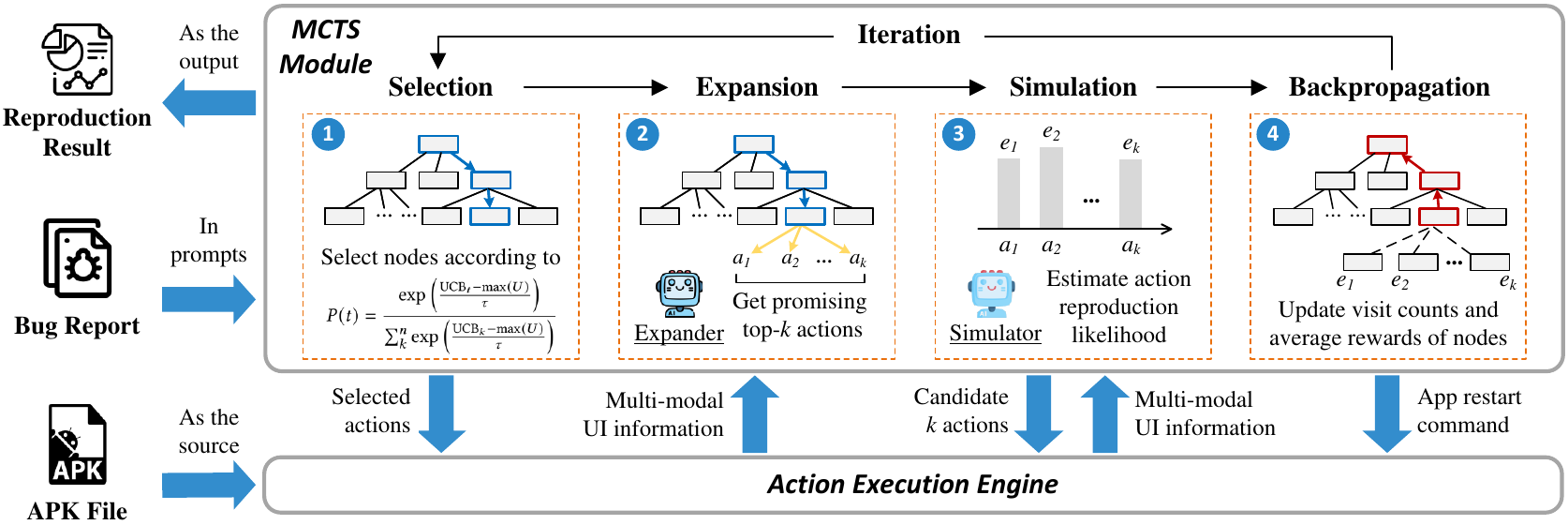}
    \caption{Overall architecture of \oursystem}
    \label{fig:arch}
\end{figure}

\section{Methodology} 
\subsection{Architecture}
\Cref{fig:arch} depicts the overall architecture of \oursystem.
It is end-to-end, requiring users to provide a bug report and an APK file.
As a result, analysts without any knowledge of LLMs can conveniently use the tool.
The final output is an action sequence that precisely reproduces the reported bug, or a message indicating that no bug is found.
\oursystem consists of two main components:

\textbf{(1) MCTS Module} guides the searching of bug-triggering action sequences by orchestrating the four stages of MCTS, \ie, selection, expansion, simulation, and backpropagation.
To balance exploitation and exploration during bug reproduction, we adapt the four stages around an LLM-based multi-agent framework.
Particularly, the two agents named Expander and Simulator respectively leverage their analytical and reasoning capabilities during the expansion and simulation stages.
This enables them to understand the app's runtime UI states and the context of bug reproduction.
Moreover, task-specific enhancements are introduced to improve computation strategies in both selection and backpropagation.
The module interacts with the Action Execution Engine to collect various types of UI-related information that support the agents' decision-making process.

    
\textbf{(2) Action Execution Engine} is an Android automation backend that enables programmatic interaction with input apps by querying and manipulating UI elements for our reproduction tasks, \eg, screen traversal and event injection.
Specifically, this component executes various commands translated from LLMs' responses in the MCTS Module, and delivers multi-modal UI information (\ie, textual descriptions and screenshots) to the LLM-guided agents of the module above.

\subsection{LLM-empowered Monte Carlo Tree Search}

We integrate an LLM-based multi-agent framework with MCTS to enable effective searching in the large and complex state space of bug reproduction.
The state space is modeled as a Monte Carlo search tree, where each node represents a state during reproduction and each edge corresponds to an input action that triggers the transition between the states.
Here, a state refers to a snapshot of an app's UI at a specific point during its execution, including the screen layout, visible elements, and their interactive properties.
This integration forms a closed-loop and feedback-driven process, where MCTS drives strategic and goal-directed searching, while LLM-guided agents provide high-level semantic understanding of the reproduction context to support the decision-making of MCTS.
By leveraging incremental feedback, the agents help MCTS adjust its searching strategy dynamically.


To operationalize this integration, we introduce LLM-guided agents into the expansion and simulation stages of MCTS, where high-level semantic reasoning and adaptive decision-making are essential.
Moreover, the selection and backpropagation, which primarily involve numerical computations, are implemented without LLMs.
Instead, we apply task-specific adaption for these two stages to better support the requirements of bug reproduction.

\subsubsection{Selection}
Starting from the root node of a search tree representing the initial UI state, we follow a path by recursively selecting child nodes based on a selection policy that balances exploration and exploitation.
In our MCTS-based framework, rewards derived from LLMs can be imperfect due to their approximate and context-sensitive nature, which may cause the standard Upper Confidence Bound (UCB)-based node selection~\cite{auer2002nonstochastic} to prematurely favor suboptimal actions and insufficiently visit alternative paths.
To address this, we adopt a variant known as the softmax-over-UCB strategy~\cite{2021FOP}, which assigns selection probabilities to available child nodes based on their UCB scores.
This approach transforms deterministic UCB scores into a softmax-based probability distribution, enabling stochastic yet guided selection that improves state space coverage and mitigates premature convergence~\cite{danihelka2022policy}.
At each selection stage, \oursystem recursively traverses the search tree from the root, and samples expanded child nodes based on softmax-normalized UCB scores, until it reaches a leaf or unexpanded node.

For a given child node $i$, its UCB score is calculated as:

\begin{equation}
    UCB_{i} = X_{i} + \sqrt{2} \cdot \sqrt{\frac{\ln N}{n_{i}}}
    \label{eq:ucb}
\end{equation}
, where $X_{i}$ denotes the average reward of node $i$, $n_{i}$ is the number of times node $i$ has been visited, $N$ refers to the number of times the parent node has been visited, $\sqrt{2}$ is a constant chosen in accordance with theoretical regret bounds derived using Hoeffding's inequality~\cite{auer2002finite}.

Given a set of child nodes with UCB scores, \ie, $U$ = $\{\text{UCB}_1, \text{UCB}_2, ..., \text{UCB}_n\}$, the selection probability for the $t$-th node is computed as:


\begin{equation}
P(t) = \frac{\exp\left(\frac{\text{UCB}_t - \max(U)}{\tau}\right)}{\sum_{k=1}^n \exp\left(\frac{\text{UCB}_k - \max(U)}{\tau}\right)}
\end{equation}
, where $\tau$ is a temperature parameter.
It controls the smoothness of the resulting probability distribution: Lower $\tau$ produces sharper preferences, while higher values yield more uniform probabilities~\cite{hinton2015distilling}.
We conduct experiments with varying $\tau$ values and find that our method is relatively insensitive to this parameter.
Thus, we empirically fix $\tau$ = 1.8 as a stable default in all experiments.
Furthermore, the subtraction of $\max(U)$ is for numerical stability to avoid overflow during exponential computation.

The selected actions are sequentially fed into the Action Execution Engine for actual execution, which then waits for the next action inputs at the subsequent simulation stage.

\subsubsection{Expansion}
This stage generates candidate actions for the search tree, enabling the exploration of previously unvisited paths.
However, standard MCTS expands only a single branch per iteration, which can limit the efficiency of reproducing Android app bugs.
Given the potentially high branching factor in UI exploration, fully expanding all child nodes at each step may introduce a large number of irrelevant actions that mislead the search process.




We introduce Expander, an LLM-guided agent designed to assist the expansion stage of MCTS by selecting a moderate amount of promising actions for further UI exploration.
At each expansion step, the Expander agent queries the LLM using the ExpansionPrompt depicted in \Cref{sec:sign}.
Instead of expanding a single action per iteration or fully expanding all possible actions at once, the agent performs a top-\textit{k} expansion strategy based on the LLM's ranking of candidate actions.
Guided by semantic cues from the bug report and the reproduction context, the LLM suggests \textit{k} candidate actions most likely to progress toward the reproduction goal.
To support this, the Action Execution Engine retrieves the multi-modal UI information at the current unexpanded leaf node.

This strategy draws inspiration from beam search, a widely adopted heuristic in sequence generation and planning tasks~\cite{wang2025proprag,sudo2024contextualized}.
Unlike arbitrary pruning, it leverages semantic relevance to the reproduction goal to guide searching.
Although some potentially useful actions may be omitted, the trade-off is deliberate and controlled.
Experimental results in \Cref{sec:exp} show that this strategy leads to effective reproduction performance, suggesting that the benefits of focused and semantically guided exploration outweigh the occasional risk of missing critical actions in practice.


\subsubsection{Simulation}\label{sec:simulation}
At this stage, \oursystem estimates the potential of each expanded candidate action to contribute to a successful reproduction, providing a value signal to guide subsequent decision-making in the MCTS process.
The stage also adds new state nodes into the search tree.
Since bug reproduction often requires long sequences of UI interactions and each step in the sequence presents numerous possible actions, this leads to a combinatorial explosion in the number of candidate simulated paths.
Moreover, unlike playing Go~\cite{silver2016mastering,silver2017mastering}, operating an app's UI in the real environment to find the termination state that triggers the reported bug faces two difficulties: (i) The search space is vast and bug-triggering states are sparse, limiting the feedback available within a finite number of interactions, and (ii) executing certain UI actions incurs high latency due to animations, loading delays, and system response times.
As a consequence, exhaustive strategies like standard Monte Carlo rollouts are computationally infeasible within a practical time budget.


We design an LLM-guided agent named Simulator, which performs lightweight simulation by estimating the potential of candidate actions to the bug reproduction goal, without executing full interaction paths.
The Simulator agent queries the LLM using the SimulatePrompt that will be also presented in \Cref{sec:sign}, following the LLM-as-a-Judge paradigm, which has been shown effective in numerous tasks~\cite{gu2024survey}.
This paradigm enables context-aware and semantically grounded evaluations without requiring task-specific supervision.

For each of the action, the Simulator agent executes it in the runtime environment using the Action Execution Engine (\ie, one-step look-ahead rollout), and observes the resulting UI transitions.
For each UI transition, the agent invokes the LLM to assess how closely it aligns with the expected behaviors described in the bug report and assign a heuristic score.
This assessment also takes into account the reproduction context accumulated along the current path in the search tree.
The output score is constrained to the range [0,10] through prompt engineering, as described in \Cref{sec:sign}.


\begin{table}[t]
\centering
\caption{Prompt examples for two distinct roles}\label{tab:prompt1}
\resizebox{\textwidth}{!}{
\begin{tabular}{m{3cm}|p{7cm}|p{7cm}}
\toprule
\rowcolor{gray!20}
\textbf{Component} & \textbf{ExpandPrompt} & \textbf{SimulatePrompt} \\ \midrule

Task specification & 
\textbf{Movitation:} 

\textit{You are an expert Android UI tester assisting me in reproducing bug reports on an emulator...You need to provide me with three suggested actions that have the highest likelihood of reproducing the bug report.} 

\textbf{Input specification:}

\textit{I will provide the app name, the bug report text, a description of the current screen layout and state and the srceenshot of the current UI page...and we will repeat this cycle until the bug reproduces...}

\textbf{Output restriction:}

\textit{You need to provide me with k suggested actions that have the highest likelihood of reproducing the bug report...Your suggestions should be formatted as follows: [\{"action": "x1", "feature": "y1"\}, \{"action": "x2", "feature": "y2"\}, ...].}
&
\textbf{Motivation:}

\textit{Suppose you are an expert in reproducing Android bugs, and your goal is to assess whether a specific action — referred to as the target action — is a valid and appropriate action toward completing the bug report, and then assign a score to that action...}

\textbf{Input specification:}

\textit{I will provide you with the bug report I am working to reproduce, along with the reproduction path I have followed so far and the corresponding UI information for each action in that path. After that, I will give you the target action...}

\textbf{Output restriction:}

\textit{Your response format must be as follows: Score: 0–10. For example, if you intend to return a score of 5, your reply should contain exactly 'Score: 5'...}
\\ \midrule

Few-shot learning &
\textbf{Example 1: }

\textbf{App name:} \textit{Alarmio} 

\textbf{Bug report:} \textit{The app gets crashes when I click about \& OSS licenses}

\textbf{UI Information}: 

\textbf{Activity:} \textit{.activities.MainActivity.} 

\textbf{Groups:} \textit{["Alarms"] , ["Settings"], ...}

\textbf{History:} \textit{Action Sequence + Current UI Screenshot }

\textbf{Reasoning: }

\textit{The bug report explicitly instructs to click "about \& OSS licenses". However, scanning the current UI groups, this specific widget is NOT present on the screen. We cannot execute the trigger directly.
Based on standard Android design patterns, "About" and "OSS Licenses" information is almost always nested inside the Settings menu.}

\textbf{Suggestions: }

\textit{1. \{'action': 'click', 'feature': 'SETTINGS'\}}

\textit{2. \{'action': 'click', 'feature': '+'\}}   

\textit{3. \{"action":"scroll","feature":"down"\} } &
\textbf{Example 1:} 

\textbf{Bug report:} \textit{When trying to log in, pressing the "Sign In" button does nothing.}

\textbf{Previous page:} \textit{Showing a login screen with email and password fields and a "Sign In" button. }

\textbf{History:} \textit{Action sequence + Current UI screenshot}

\textbf{Target action:} \textit{Click the "Sign In" button.} 

\textbf{Result:} \textit{The button is clearly relevant to the bug report and is the correct next action.}

\textbf{Score:} \textit{9}\par\medskip\noindent

\textbf{Example 2: }

\textbf{Bug report:} \textit{When scrolling down in the article list, some items disappear.}

\textbf{Previous page:} \textit{Article list screen.}

\textbf{Target action:} \textit{Click the "Back" button.} 

\textbf{History:} \textit{Action sequence + Current UI screenshot}

\textbf{Result:} \textit{This action does not move toward reproducing the scrolling bug.} 

\textbf{Score:} \textit{2} \\ \midrule

Chain-of-thought &
\textbf{When generating suggestions:} 

\textit{1. Identify if the target widget mentioned in the bug report is directly visible in the current UI. }

\textit{2. If visible → Suggest clicking it as first priority}. 

\textit{3. If not visible → Decide if scrolling is likely to reveal it}. 

\textit{4. If still missing → Systematically explore all first-level widgets. }

\textit{...}&
\textbf{When scoring:} 

\textit{1. Read the bug report carefully to identify the core triggering action or condition.} 

\textit{2. Compare the target action with the bug report goal: Does it directly interact with the suspected element or trigger? Does it logically advance toward reproducing the bug?} 

\textit{3. Check the UI change (before vs after action)...}

\textit{4. Assign score...} \\ 

\bottomrule
\end{tabular}
}
\end{table}

To ensure independent simulation of each action, \oursystem employs a UI state recovery mechanism that restores the app to the parent node's state after each simulation.
It records the selected actions during expansion and thus maintains the full action sequence from the root to each expanded node.
This enables \oursystem to restart the app and replay the recorded actions to accurately reconstruct the corresponding UI state.
The mechanism is implemented by sending the corresponding commands to the Action Execution Engine.



\subsubsection{Backpropagation}\label{subsec:bak}

\oursystem leverages the outcomes of the simulation stage to update the statistics of nodes along the current search path.
Specifically, the visit count of each node is incremented by 1, and the average reward of each node is updated using the mean of the simulation scores of the \textit{k} newly-expanded nodes.
These updates refine the UCB scores of the corresponding nodes, directly influencing action selection in the subsequent MCTS process.

After backpropagating the score up the tree, the UI state of the target app is restored to the root node of the search tree using the recovery mechanism introduced in \Cref{sec:simulation}, preparing it for the next iteration of MCTS.

\subsection{Prompt Design}\label{sec:sign}
Prompt engineering plays a crucial role in enabling LLM-guided agents to understand contextual information and effectively reproduce bugs in Android apps.
Existing techniques have proposed various prompts for bug reproduction~\cite{wang2024feedback,feng2024prompting,huang2025one}.
However, these prompts cannot be directly applied to our work, since they are not designed to integrate with external planning strategies (\eg, MCTS), which are crucial for achieving the UI state space exploration.

To overcome the limitation, we design two sets of role-specific prompts, \ie, ExpandPrompt and SimulatePrompt listed in \Cref{tab:prompt1}, to enable seamless integration of general-purpose LLMs into our workflow.
Note that the prompts shown in the table are simplified versions. In real-world scenarios, they are typically longer and include more contextual information.
Specifically, these prompts instantiate LLMs as domain-aware agents, aligned with the expansion and simulation stages of the MCTS process respectively.
To improve the agents' ability to perform complex tasks and support downstream decision-making, we apply advanced prompt engineering techniques, namely few-shot learning~\cite{liu2023pre} and chain-of-thought reasoning~\cite{wei2022chain}.
These techniques not only enable LLMs to perform modular task decomposition and multi-step reasoning, but also provide structured instructions and ensure consistent output formatting.

\subsubsection{Task Specification}
While LLMs possess general reasoning capabilities, they often lack the task-specific focus required for bug reproduction.
We design role-specific instructions that specialize LLMs into dedicated agents. 
As listed in the second row of \Cref{tab:prompt1}, we use tailored prompting strategies for each agent's responsibilities, enabling general LLMs to emulate the behavior of different experts for automated bug reproduction.

The motivation prompts are designed to clearly define the LLMs' roles and objectives, guiding its behavior in our task.
They explicitly instruct LLMs not only to execute the reproduction steps, but also to observe and confirm whether the specific buggy behavior described in the input report is actually triggered.
This goal enhances both the reliability and accuracy of the reproduction process.

To help the LLM-guided agents interpret the environment, we provide well-structured input specifications that encode both textual and visual information about the app and its UI.
Specifically, the input includes the app name, the textual bug report, the structured description of the visited UI components, and the current UI screenshot.
The descriptive UI information extracted from XML files of app pages provides structured and semantic details (\eg, widget hierarchy), whereas the screenshots capture visual and spatial context. Together, they offer complementary perspectives to better understand the screen's functional layouts and their spatial relationships.

To enable deterministic parsing, we define a structured output format for each prompt type. 
This guarantees consistent communication with downstream components and significantly reduces integration errors from verbose or inconsistent outputs.
In the ExpandPrompt, LLMs are required to generate \textit{k} most probable actions for successful bug reproduction and calculate their coordinates in a single response, adhering to the strict JSON-like format (\eg, \textit{\{"action": "x1", "feature": "y1"\}}), as illustrated in \Cref{tab:prompt1}.
Inspired by the action format in ReBL, we restrict the output to six predefined actions: \textit{click}, \textit{long\_click}, \textit{set\_text}, \textit{multiple\_select}, \textit{rotate}, and \textit{back}.
This prompt-based action generation format is used during the expansion stage of our MCTS process.
For the SimulatePrompt, LLMs are prompted to output a numerical score in a fixed format.
This ensures compatibility with the backpropagation stage in MCTS.




\subsubsection{Few-shot Learning}\label{subsub:fewshot}
A representative example helps the LLM elicit specific knowledge and abstractions needed to complete the downstream task.
To construct the few-shot examples for guiding the LLM, we follow the methodology in AdbGPT~\cite{feng2024prompting} and invite five experienced participants to help identify challenging cases.
Specifically, three developers collaboratively construct representative examples that guide the model in handling realistic and non-trivial scenarios, and highlight its reasoning process, especially when bug reports lack key reproduction steps.
This construction is grounded in real-world bug reports from our dataset.
To further ensure quality and domain relevance, we also invite two professors with expertise in Android crash reproduction to join the process. All five participants engage in extensive discussions and ultimately reach a consensus on the final set of eight representative examples used during reproduction, four pertaining to ExpandPrompt and four to SimulatePrompt.

As shown in the third row and and second column of \Cref{tab:prompt1}, the ExpandPrompt enables LLMs to learn real-world crash reproduction patterns via few-shot demonstrations.
It teaches the LLM not only to generate promising actions in a structured format, but also to reason about missing steps.
When a UI element mentioned in the bug report is not present on the current page, the ExpandPrompt guides the LLM to infer the most plausible action to reach the intended UI page.
For example, in the few-shot instance from \Cref{tab:prompt1}, the target element \textit{about \& OSS licenses} is missing.
Through demonstration, the LLM learns to generate the action \textit{click "Settings"} instead, based on its relevance to the bug report.

In the SimulatePrompt, the examples illustrate how to score a target action based on the current UI state and exploration history.
The third row and third column of \Cref{tab:prompt1} lists two contrasting examples that highlight the difference between high- and low-scoring actions.
In the first example, the bug report exactly matches the target action, indicating that it should receive a very high score.
Conversely, the second example presents a target action that likely deviates from the correct reproduction sequence, and should therefore receive a low score.


\subsubsection{Chain-of-thought Reasoning}
We adopt chain-of-thought (CoT) reasoning to encourage step-by-step thinking, helping LLMs articulate intermediate steps explicitly, resolve ambiguity, and produce more consistent and interpretable outputs in bug reproduction scenarios.
Here, we also follow the methodology in AdbGPT~\cite{feng2024prompting} by asking the three developers mentioned in \Cref{subsub:fewshot} to provide chain-of-thought reasoning for our UI exploration task.

As shown in the fourth row of \Cref{tab:prompt1}, CoT reasoning is used to guide LLMs through intermediate reasoning steps in domain-specific bug reproduction cases. 
For example, in the ExpandPrompt, we distill a set of reasoning strategies to help LLMs generate and evaluate candidate actions, increasing the likelihood of successful reproduction. 
In the SimulatePrompt, we provide structured and step-by-step reasoning guidance for LLMs to assess and score target actions reliably.

\section{Experimental Evaluation}\label{sec:exp}

To evaluate the effectiveness of \oursystem, we seek to answer the following three questions:

\begin{itemize}
    \item \textbf{RQ1:} How does \oursystem compare to state-of-the-art methods in bug report reproduction?
    \item \textbf{RQ2:} How do each proposed strategy contribute to the overall effectiveness of TreeMind?
    \item \textbf{RQ3:} What is the optimal \textit{k} for the expansion stage in MCTS?
\end{itemize}

\subsection{Experimental Setup}
\subsubsection{Implementation}
We implement a prototype of \oursystem in Python.
Specifically, we use an Android virtual device to install and run apps, and employ UI Automator2~\cite{uiautomator2} to drive their execution according to the input actions.
We then extract the textual information about apps' UI using the off-the-shelf implementation of ReBL.
Besides, we leverage the GPT-4o~\cite{gpt} model\footnote{Version: gpt-4o-2025-03-26} provided by OpenAI to support the understanding and reasoning over textual descriptions and images.
Following the configuration in ReBL, we set the temperature of the LLM to 0.3 to reduce randomness and ensure more deterministic outputs.

\subsubsection{Dataset}
To avoid potential bias and ensure representativeness, we collect 93 real-world textual crash reports and the relevant APKs from three existing open-source datasets released by ReActDroid (22 out of 76) ~\cite{huang2025one}, ReproBot (66 out of 76) ~\cite{zhang2023automatically}, and AndroR2 (5 out of 90)~\cite{johnson2022empirical} respectively.
We refine the original datasets by excluding duplicate reports and those associated with non-installable APK files.
Specifically, duplicate reports include 4 between AndroR2 and ReproBot, 22 within ReActDroid, and 5 between AndroR2 and ReActDroid.
Installation failures include 81 in AndroR2, 27 in ReActDroid, and 10 in ReproBot.



\subsubsection{Baselines}\label{sec:baseline} To evaluate the effectiveness of \oursystem, we select four state-of-the-art methods for reproducing bug reports of Android apps: 
\begin{itemize}
    \item ReBL~\cite{wang2024feedback} is a feedback-driven technique that uses the complete textual bug report and innovative prompts to automatically reproduce bugs of Android apps.
    \item ReActDroid~\cite{huang2025one} is an LLM-based approach for reproducing crashes of Android apps from one-sentence overviews, by leveraging novel prompts to derive exploration steps.
    \item AdbGPT~\cite{feng2024prompting} is the first work to utilize prompt engineering with few-shot learning and chain-of-thought reasoning to harness LLMs' knowledge for automated bug replay.
    \item ReproBot~\cite{zhang2023automatically} designs a NLP-based analysis to extract reproduction steps from Android bug reports, and finds the match between steps and UI events by reinforcement learning (RL).
\end{itemize}

Note that we use the same version of GPT-4o across all three LLM-based tools to ensure fairness.

\subsection{RQ1: Effectiveness and Efficiency of \oursystem}\label{sec:RQ1}
To compare the effectiveness and efficiency of our approach against the baselines, we test each collected bug report along with its corresponding APK on \oursystem and each selected baseline.
This process is repeated five times for each technique, with a time limit of 30 minutes per run.
A bug is considered successfully reproduced if at least one of the five runs succeeds.
Among the successful attempts, we record the shortest reproduction time as the final time cost.

\begin{table}[t]
\centering
\caption{Comparison of \oursystem and baselines on bug reproduction ( RCRs = Reproduced Complete Reports, RIRs = Reproduced Incomplete Reports)}
\label{tab:baseline}
\newcolumntype{C}[1]{>{\centering\arraybackslash}m{#1}}
\resizebox{\textwidth}{!}{
\renewcommand{\arraystretch}{1.1} 
\setlength{\tabcolsep}{6pt}       
\begin{tabular}{C{2.5cm}|C{2cm}|C{2cm}|C{2cm}|C{2cm}|C{2cm}}
\toprule
\rowcolor{gray!20}
\textbf{Metric} & \textbf{\oursystem} & \textbf{ReActDroid} & \textbf{ReBL} & \textbf{AdbGPT} & \textbf{ReproBot} \\
\midrule
\textbf{Success rate (\%)} & \textbf{64.52} & 	45.16 & 40.86 & 34.41 & 31.18 \\
\textbf{\# of RCRs} & \textbf{30} & 21 & 23 & 20 & 16 \\
\textbf{\# of RIRs} & \textbf{30} & 21 & 15 & 12 & 13 \\
\textbf{Time cost (s)} & 241.4 & \textbf{87.5} & 123.4 & 152.7 & 827.2 \\
\textbf{Token usage} & 183,163 & \textbf{125,548} & 158,245 & 138,641 & - \\
\bottomrule
\end{tabular}
}
\end{table}

\subsubsection{Effectiveness}\label{sec:eff}
The second row of \Cref{tab:baseline} presents the success rate of reproducing bug reports from our collected dataset.
Overall, \oursystem reproduces 64.52\% of the reports (60 out of 93), significantly outperforming the baslines: ReActDroid by 42.9\%, ReBL by 57.9\%, AdbGPT by 87.5\%, and ReproBot by 106.9\%.
Moreover, \oursystem successfully reproduces 30 RCRs and 30 RIRs.
In comparison, ReActDroid, ReBL, AdbGPT, and ReproBot reproduced 21/21, 23/15, 20/12, and 16/13 RCRs/RIRs, respectively.
Here, a report is considered incomplete if, at any UI state, it fails to specify an unambiguous action that leads to the next UI state.
We manually inspect each report in our dataset and record the counts of RCRs and RIRs above.

The superior performance of \oursystem can be mainly attributed to its integration of an external planning strategy (\ie, MCTS) with LLM semantic reasoning.
LLMs excel at interpreting both textual and visual information to generate context-aware actions, while MCTS guides the searching toward more promising directions.
This integration enables \oursystem to reproduce bugs more effectively, even when given incomplete reports.
In contrast, the techniques that relying solely on LLMs~\cite{wang2024feedback,huang2025one,feng2024prompting} are primarily reactive and may struggle with long-horizon planning.
The RL-based approach~\cite{zhang2023automatically} faces difficulties when handling tasks that require semantic understanding, cross-task generalization, and adaptive reasoning.

\noindent\textbf{Analysis of Failure Cases.} While \oursystem performs well in most cases, some crashes remain difficult to reproduce due to limitations that are common across automated tools~\cite{wang2024feedback,huang2025one}.
Among the 33 failed cases, we identify three major causes: (i) missing critical details in the bug report (27 cases), (ii) third-party service dependencies (5 cases), and (iii) UI Automator2 limitations (1 case).

First, incomplete bug reports often hinder effective UI exploration.
For instance, in the app named \textit{Markor-1565}, the report states that the app crashes "\textit{after rotating the note to landscape or back to portrait}".
\oursystem runs for over 30 minutes without success.
Manual attempts also fail, suggesting that key interactions are missing or not clearly specified.



Second, as noted in ReBL, crashes involving third-party services pose challenges for automated tools.
In the app named \textit{yakusu-35}, the reproduction requires navigation to Google and selection of a specific account. 
Under such circumstances, \oursystem fails to reproduce the crash.

Third, the limitations of UI Automator2 affect both UI extraction and action execution, as also noted in ReBL.
In the app named \textit{Memento-169}, the tool fails to extract custom views from the UI hierarchy and cannot accurately locate the date picker dialog, resulting in imprecise clicks instead of accurately scrolling to the targeted item.
Although \oursystem successfully triggers the crash, further analysis shows that it stems from an unintended interaction: The appearance of the on-screen keyboard shifts the UI layout, causing a misclick that inadvertently reproduces the bug.

\noindent\textbf{Comparative Case Analysis.} To further demonstrate effectiveness, we compare \oursystem with ReActDroid and ReBL on representative cases.
We identify three instances where \oursystem fails but ReActDroid succeeds.
The failures are caused by incorrect path scoring during early exploration (\ie, \textit{yakusu-1}), excessive search time due to long reproduction paths (\ie, \textit{saner2022-1299}), and difficulties in handling a multi-step login interface (\ie, \textit{saner2022-129}).
In contrast, ReActDroid successfully handles all three cases, benefiting from GUI pre-processing and well-tailored prompts.
Furthermore, we identify two instances (\ie, \textit{saner2022-23} and \textit{saner2022-271}) where \oursystem fails while ReBL succeeds. 
These failures stem from \oursystem's use of MCTS-based planning, during which certain UI pages appear only once in the entire app execution.
Since these pages are essential for reproducing the bugs but cannot be revisited, the reproduction ultimately fails.
Reinstalling the apps after each MCTS iteration could resolve these issues, but would significantly increase the overall runtime.


Conversely, \oursystem successfully reproduces bugs in 20 and 24 cases where ReActDroid and ReBL fail, respectively.
These successes are largely attributed to our LLM-empowered MCTS, which effectively narrows down the search space and progressively guides the tool toward the correct sequence of actions, even in the absence of critical steps in the crash report.
As shown in \Cref{fig:case}, where ReActDroid and ReBL repeatedly rotate the screen until timeout, whereas \oursystem, guided by MCTS and the LLM, identifies the correct reproduction path within a few iterations.
In the case of \textit{recdroid-56}, although the bug report contains a complete sequence of actions, both ReBL and ReActDroid fail to reproduce the crash. In contrast, \oursystem successfully reproduces the bug.
Our success is attributed  to the MCTS iterations, during which the context leading to the bug is reconstructed.
We elaborate on this process in \Cref{sec:casestudy} below.

These results suggest that combining the strengths of \oursystem and other advanced tools (\eg, ReActDroid and ReBL) has the potential to improve automated bug reproduction.

\subsubsection{Efficiency}
As shown in the third row of \Cref{tab:baseline}, our tool takes an average of 241.4 seconds per successful reproduction, with the time ranging from 13 to 1095 seconds depending on the specific bug. 
On average, the tool queries the LLM approximately 18.75 times per bug.
This time consumption is relatively higher than that of the LLM-based baseline tools, primarily due to the multi-round planning process adopted by \oursystem.
ReActDroid demonstrates the shortest execution time among all the tools mentioned above, with only 87.5 seconds.
ReproBot takes the longest time, 827.2 seconds, which may be attributed to both the S2R matching process and the convergence time required by the RL algorithm.

Although \oursystem incurs higher time costs, its ability to reproduce a wider range of bugs justifies the efficiency trade-off, particularly in scenarios where reliability and reproduction success take precedence over speed.
When considering the total time spent across all test cases, regardless of timeout or success, \oursystem proves to be the fastest overall due to the highest success rate.

In terms of token usage and dollar cost, \oursystem consumes the most tokens (183,163) with an estimated cost of \$0.45 per successfully reproduced bug.
This is primarily due to the longer time required to successfully reproduce a bug, and each MCTS iteration involves both textual and visual inputs to the LLM.
ReActDroid is the most cost-efficient, using 125,548 tokens (\$0.22), followed by ReBL (158,245 tokens, \$0.25) and AdbGPT (138,641 tokens, \$0.24). 
ReproBot does not rely on LLMs and thus incurs no token-related cost.



\begin{figure}[t]
    \centering
    \includegraphics[width=\textwidth]{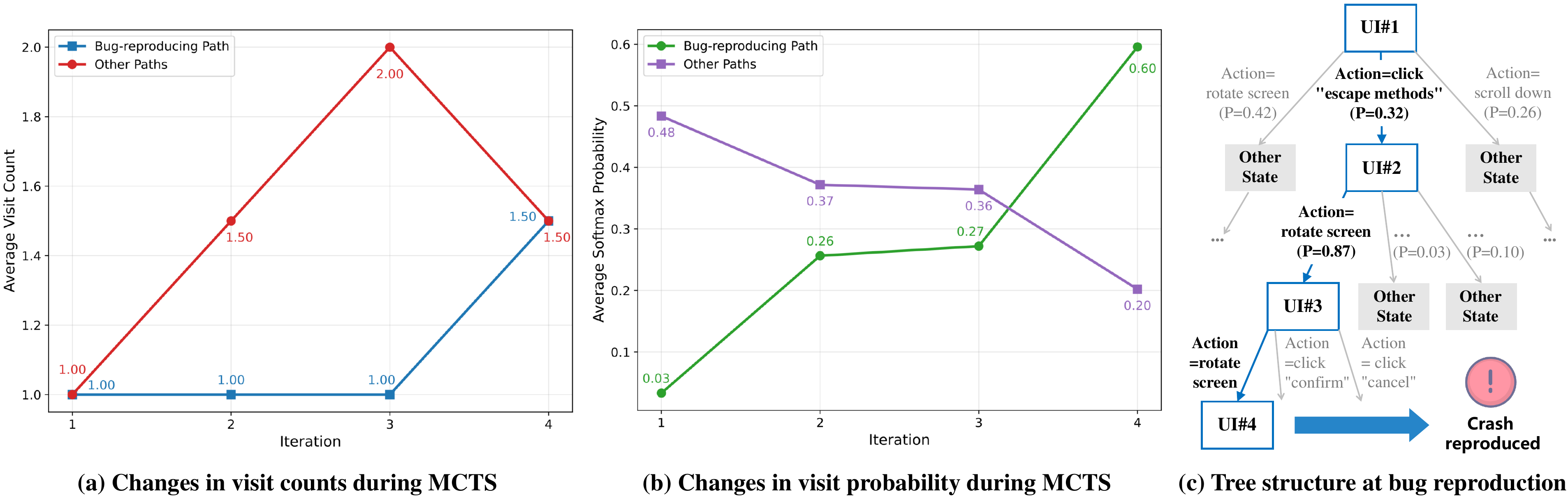}
    \caption{MCTS Statistics for reproducing the real-world bug shown in \Cref{fig:case} using \oursystem}
    \label{fig:cs}
\end{figure}

\subsubsection{Real-world Case Studies}\label{sec:casestudy}
We select two representative cases to demonstrate the advantages of \oursystem.
One of the bug reports omits essential actions needed to reproduce the bug (\ie, \textit{Case 1}), while the other includes a complete and accurate sequence of actions (\ie, \textit{Case 2}).


\noindent\textbf{Case 1.} In our experiment, none of the baselines described in \Cref{sec:baseline} are able to solve the real-world case presented in \Cref{sec:motivation} within 30 minutes.
By contrast, \oursystem provides an effective and systematic approach for reproducing the crash triggered by screen rotation.
The entire reproduction process takes 3 minutes and 41 seconds.
Since MCTS is an iterative process involving multiple rounds of updates, as shown in \Cref{fig:cs}, we show how \oursystem solves the case by analyzing the details stored in the search tree.
Note that the figure illustrates only one possible way in which \oursystem reproduces the bug. 
In practice, \oursystem can reproduce the bug through different action sequences.

As shown in \Cref{fig:cs}(a), during the first three iterations of MCTS, the average visit count of nodes along the bug-reproducing path remains low and unchanged, while other nodes are visited more frequently, peaking at iteration 3.
In iteration 4, the average visit count for the bug-reproducing path increases sharply, whereas that of the other nodes declines.
This suggests that the search initially explores suboptimal paths, but gradually shifts focus toward the correct path as more information becomes available.

As shown in \Cref{fig:cs}(b), the average softmax probability of the bug-reproducing path starts very low but increases steadily across MCTS iterations, reaching 0.60 at iteration 4. In contrast, the probability assigned to other paths consistently decreases from 0.48 to 0.20. This trend indicates that the search gradually learns to assign higher confidence to the correct path, improving its ability to prioritize promising actions over time.
In other words, actions on the bug-reproducing path initially have low UCB scores and visit probabilities.
Instead of greedily selecting the highest-UCB action, \oursystem applies softmax-over-UCB sampling to explore low-probability actions and eventually converge to the correct path.

As depicted in \Cref{fig:cs}(c), \oursystem successfully triggers the bug along the blue-highlighted path.
The reproduction process starts from the app's initial screen (\ie, UI\#1), where three candidate actions are expanded: \textit{rotate screen}, \textit{click "escape methods"}, and \textit{scroll down}.
\oursystem selects the action \textit{click "escape methods"}, transitioning to UI\#2 with a probability of 32\%.
From there, it executes \textit{rotate screen}, leading to UI\#3 with a probability of 87\%.
Finally, another \textit{rotate screen} action leads to UI\#4, where the crash is reproduced.
Note that the transition probability from UI\#3 to UI\#4 is not shown, as this action is directly executed during the expansion stage, rather than being sampled via the softmax-over-UCB strategy at the selection stage.

This case shows the advantage of combining MCTS's exploratory planning with LLM's semantic reasoning. 
Even when critical steps are missing from the report, our tool can effectively recover them and reconstruct the complete crash-inducing sequence.
Although other baseline tools do not reproduce this bug within 30 minutes in our experiments, this may be due to factors such as the randomness of LLM outputs. 
With more attempts or extended time, they might still succeed.


\begin{figure}[t]
    \centering
    \includegraphics[width=\textwidth]{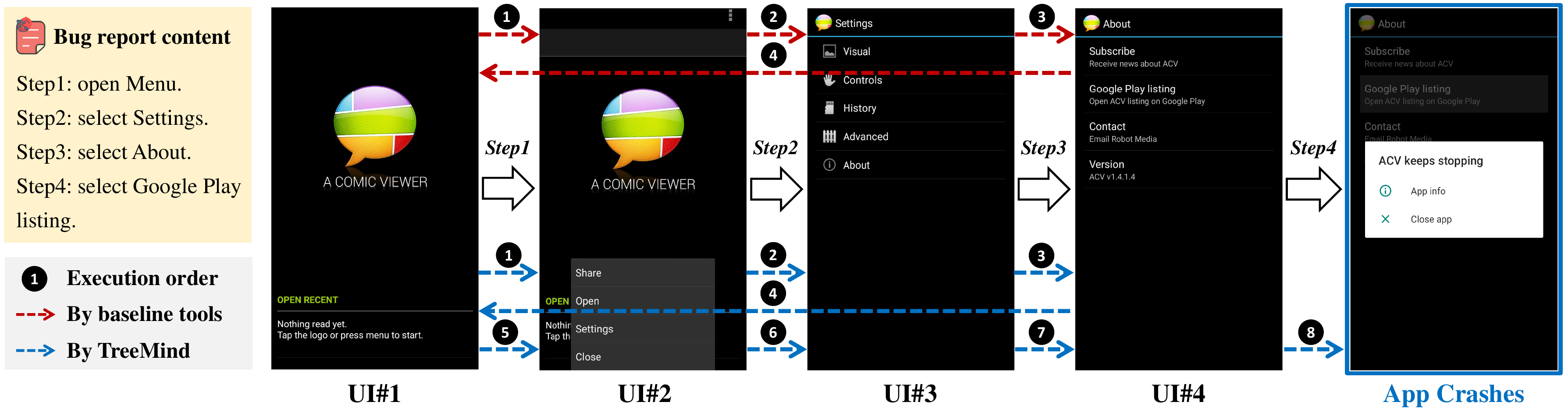}
    \caption{\oursystem vs. baseline tools: A representative case with complete action sequences from the bug report}
    \label{fig:Case1}
\end{figure}

\noindent\textbf{Case 2.}
As explained in \Cref{sec:eff}, the case of \textit{recdroid-56} is not successfully reproduced by ReActDroid and ReBL within 30 minutes in our experiment.
Upon closer inspection, we find that although the bug report provides a complete sequence of actions to be executed, it omits critical contextual information required for reproduction.
Specifically, this bug is triggered only when the same sequence of actions is executed multiple times, making it different from typical cases where a single execution is sufficient to reproduce crashes.

As depicted in \Cref{fig:Case1}, the bug can be reproducible by following the hollow arrows in the center, which correspond directly to the steps listed in the top-left bug report.
However, the baseline tools do not trigger the crash after selecting \textit{Google Play listing} at UI\#4.
Instead, the app returns to UI\#1 along the red dashed arrows.
Since these tools have already executed the actions described in the report once without observing a crash, they tend to stop further exploration along this path and instead switch to exploring alternative paths.

In contrast, \oursystem employs an iterative MCTS process that naturally re-executes promising action sequences across iterations (\ie, the blue dashed arrows).
Specifically, in an earlier MCTS iteration, our tool first executes Steps 1 through 4 as described in the bug report, driving the app to UI\#3 and then back to UI\#1.
In the subsequent MCTS iteration, our tool restarts from UI\#1 and re-executes Steps 1 through 4 to continue the UI exploration, successfully triggering the bug.
Interestingly, the repeated execution reconstructs the required app context.


\begin{table}[t]
\centering
\caption{Ablation study of \oursystem}
\label{tab:abl}
\newcolumntype{C}[1]{>{\centering\arraybackslash}m{#1}}
\resizebox{\textwidth}{!}{
\renewcommand{\arraystretch}{1.1} 
\setlength{\tabcolsep}{6pt}       
\begin{tabular}{C{2.5cm}|C{2cm}|C{2cm}|C{2cm}|C{2cm}|C{2cm}}
\toprule
\rowcolor{gray!20}
\textbf{Metric} & \textbf{Original} & \textbf{\textit{w/o} PE} & \textbf{\textit{w/o} IG} & \textbf{\textit{w/o} TK} & \textbf{\textit{w/o} SI} \\
\midrule
\textbf{Success rate (\%)} & \textbf{64.52} & 47.31 &  45.16 & 26.88 & 36.56 \\
\textbf{Time cost (s)} & \textbf{241.4} & 429.0 & 345.6 & 323.4 & 335.9 \\
\bottomrule
\end{tabular}
}
\end{table}

\subsection{RQ2: Contribution of Each Proposed Strategy}\label{sec:ablation}
We conduct ablation studies to systematically evaluate the impact of individual strategy on the effectiveness and efficiency of \oursystem by comparing it against a fully functional version.
We use all bug reports in the datasets for the experiments.
Other experimental settings remain consistent with those described in \Cref{sec:RQ1}.
Specifically, we construct four ablations:
\begin{itemize}
    \item \textbf{\textit{w/o} PE} excludes the few-shot learning and chain-of-thought reasoning.
    \item \textbf{\textit{w/o} IG} does not input UI screenshots into LLMs. 
    \item \textbf{\textit{w/o} TK} outputs all possible actions without querying the LLM for top-\textit{k} candidates.
    \item \textbf{\textit{w/o} SI} adopts the standard Monte Carlo rollout strategy.
\end{itemize}

\Cref{tab:abl} presents the experimental results, demonstrating that our original version with full functions outperforms the alternative versions in both detection effectiveness and efficiency.

\begin{figure}[t]
   \centering
   \includegraphics[width=0.95\textwidth]{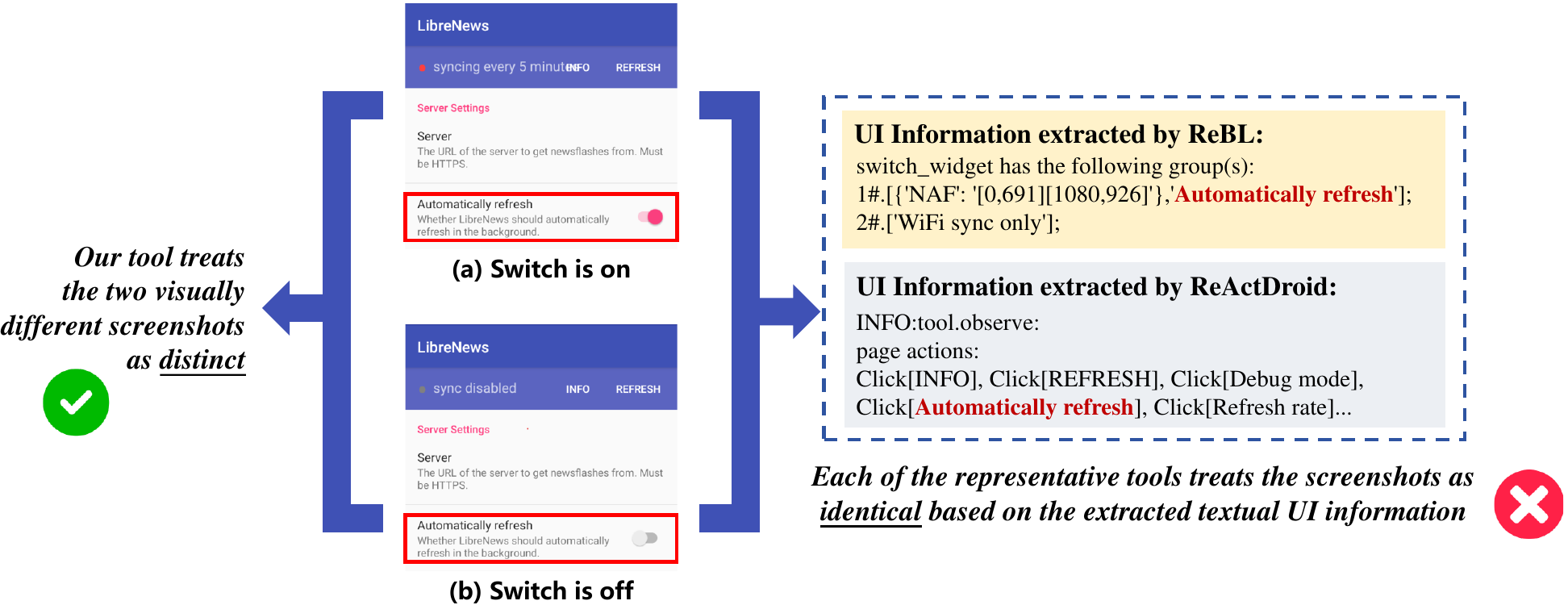}
   \caption{Comparison of UI semantic understanding via textual information and screenshots}
   \label{fig:img}
\end{figure}

\textbf{\textit{w/o} PE} achieves an overall success rate of 47.31\%, 17.21\% lower than the original version, and takes 429.0 seconds on average to reproduce a bug report, approximately 1.57$\times$ longer.
These results indicate that removing prompt engineering significantly degrades reproduction performance.
The lack of task-specific guidance and step-by-step reasoning hinders contextual understanding and leads to lower success rates and longer execution times.


\textbf{\textit{w/o} IG} achieves an overall success rate of 45.16\%, which is 19.36\% lower than the original version, and requires an average of 345.6 seconds to reproduce a bug report, approximately 1.26$\times$ longer.
These results highlight the importance of UI screenshots.
Without visual context, the tool relies solely on XML-based UI descriptions, which often lack critical visual cues.
This limitation weakens the tool's ability to understand UI semantics, resulting in reduced reproduction effectiveness.
As illustrated in \Cref{fig:img}, in the app named \textit{LibreNews}, the screenshot-disabled version of \oursystem and two representative tools (\ie, ReBL and ReActDroid) fail to distinguish whether the "\textit{Automatically refresh}" switch is on or off, because the XML-based UI information is identical across both states.
Consequently, they repeatedly click the same switch until the timeout is reached in our experiment.
In contrast, the visual difference in the screenshots allows \oursystem to identify the correct state, and thus turn off the switch once and avoid redundant actions.
While the inability to perceive a UI state does not always prevent bug reproduction for existing tools, this case shows that misinterpreting the UI can lead to incorrect behavior and ultimately reproduction failure.

\textbf{\textit{w/o} TK} achieves a 26.88\% success rate, 37.64\% lower than the original version, and takes 323.4 seconds on average to reproduce a bug, approximately 1.18$\times$ slower.
This variant exhibits the lowest success rate among all four ablation settings.
Even when some bugs are reproduced, it requires more time.
These results suggest that the absence of a limit on the number of expanded child nodes may lead to a search space explosion, which in turn hinders timely bug reproduction.


\textbf{\textit{w/o} SI} achieves an overall success rate of 36.56\%, 27.96\% lower than the original version, and takes 335.9 seconds on average for a bug report, approximately 1.23$\times$ greater.
The results demonstrate the crucial role of the optimized simulation stage.
Without it, the tool struggles to obtain accurate rewards under the standard rollout strategy, which negatively impacts the entire MCTS process.
For example, when reproducing a crash in an app named \textit{Anki-9914}, the simulation stage at this ablation assigns similar scores to different actions, making it difficult to distinguish correct from incorrect ones.
As a result, the reproduction process deviates from the correct path, undermining the intended advantage of MCTS.


\begin{figure}[t]
   \centering
   \includegraphics[width=0.6\textwidth]{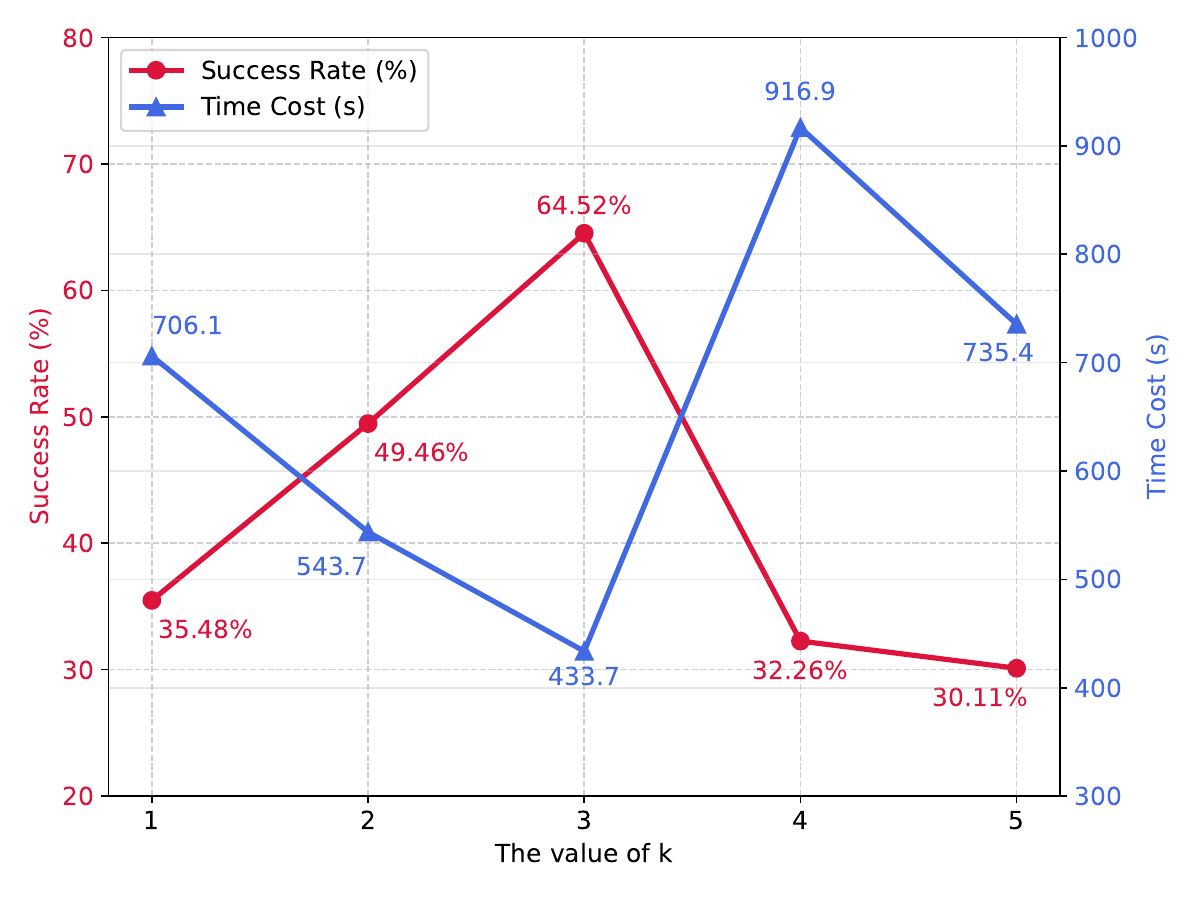}
   \caption{Success rate and time cost of \oursystem under different values of \textit{k}}
   \label{fig:topk}
\end{figure}

\subsection{RQ3: Optimal \textit{k} in MCTS Expansion}\label{sec:rq3}
To determine the optimal value of \textit{k} for our task, we conduct experiments with \textit{k} values from 1 to 5.
To ensure fairness, we use the same dataset described in \Cref{sec:ablation}.
All other experimental settings remain consistent with those described in \Cref{sec:RQ1}.

As the results in \Cref{fig:topk}, \oursystem achieves the best performance when \textit{k} = 3.
Overall, we observe that the reproduction success rate reaches its peak (\ie, 64.52\%) while the average time cost is also minimized (\ie, 241.4 seconds) under this setting.
This indicates that setting \textit{k} to 3 provides the optimal balance between exploration and exploitation in the decision-making process.

When \textit{k} < 3, although the detection success rate shows an upward trend, it remains relatively low.
Meanwhile, the average time cost stays high.
This is likely because the limited candidate action set misses key actions necessary for reproduction, which ultimately compromises decision quality.

When \textit{k} > 3, the detection success rate drops and the average time cost increases.
A larger candidate set tends to introduce more irrelevant or misleading actions, expanding the search space and complicating the decision-making process.
Interestingly, the overall time cost at \textit{k} = 5 is lower than at \textit{k} = 4.
After further analysis, we find that when \textit{k} is set to 4, \oursystem tends to focus more on reproducing complex bugs successfully, even at the cost of longer execution time.
In contrast, when \textit{k} is set to 5, it quickly reproduces simpler bugs but struggles to handle more complex ones.

\section{Discussion}
To ensure the representativeness and diversity of our evaluation data, we collect bug reports and the relevant APKs from three widely-used datasets constructed by prior studies  published in top-tier conferences and journals~\cite{huang2025one,zhang2023automatically,johnson2022empirical}.
This helps improve the external validity of our evaluation by reflecting real-world bug reproduction scenarios.
In the future, we will extend our datasets with bug reports from different domains to strengthen our work.

Our work primarily relies on two off-the-shelf techniques, including LLMs and UI Automator2, both of which have inherent limitations that may impact the performance of \oursystem.
On the one hand, LLMs may produce hallucinated or inconsistent outputs~\cite{huang2025survey}, which can undermine the effectiveness of MCTS-based planning.
To mitigate the issue, we adopt GPT-4o, a state-of-the-art LLM that supports multi-modal inputs and is widely adopted in both academia and industry.
With the rapid advancement of LLMs, we plan to explore more advanced models in the future to evaluate their effectiveness on our approach.
Moreover, the LLM training data is not public, so some bug reports might have been seen during pretraining.
However, crash reproduction involves reasoning over app-specific UI states and interactions, which goes beyond simple memorization.
To ensure the reliability of our results, we execute \oursystem five times per bug report and evaluate its performance using aggregated outputs.
On the other hand, UI Automator2, one of the most widely adopted tools for UI testing~\cite{feng2024prompting,zhang2023automatically,zhao2022recdroid+}, exhibits limitations in extracting UI information and performing certain actions, as described in \Cref{sec:eff}.
We plan to use image-based UI testing techniques~\cite{2024Mobile} to address the limitations of extracting textual UI descriptions for bug reproduction.

\oursystem achieves a high success rate for bug reproduction, but takes more execution time and tokens compared to purely LLM-based methods~\cite{wang2024feedback,huang2025one,feng2024prompting}.
This is primarily due to the integration of MCTS-based searching into the LLM-guided agents and the need to provide multi-modal UI information, which introduces additional computational overhead. 
However, we argue that this overhead is justified by the significantly improved reproduction success rate achieved by our work.
In practical scenarios, the reliability and completeness of bug reproduction are often prioritized over execution speed, especially in debugging and quality assurance workflows.
In future work, we plan to optimize the search process (\eg, via parallelization~\cite{2008Parallel}) to reduce execution time and tokens without compromising accuracy.
We also consider adopting the idea from ReBL, where the LLM outputs a sequence of actions in one round, to improve reproduction efficiency.
We believe that the above optimization strategies are orthogonal to our core approach and can be leveraged independently to further enhance performance.

\section{Related Work}

\subsection{Non-LLM-based Bug Report Reproduction}
Most existing work focuses on reproducing bug reports in textual form.
ReCDroid~\cite{zhao2019recdroid} combined natural language processing (NLP) and dynamic GUI exploration to reproduce the bugs for Android apps. ReCDroid+~\cite{zhao2022recdroid+} extended ReCDroid by leveraging deep learning techniques to extract S2R sentences.
ReproBot~\cite{zhang2023automatically} used NLP techniques to extract S2R entities and then adopted Q-learning to guide the search for successful reproducing steps.
Roam~\cite{zhang2024mobile} leveraged the target app's UI transition graph to globally search for the UI event path for reproducing bug report steps.
ScopeDroid~\cite{huang2023context} designed a multi-modal neural network that matches reproducing steps to GUI widgets by considering their icons and contextual information obtained from a state transition graph.
Traditional NLP-based methods often struggle to accurately extract S2Rs and to understand the semantics of bug reports and UI context. Moreover, the lack of global planning in some approaches may limit their effectiveness in practice.

Images are another commonly used information  modality for reproducing bug reports.
GIFdroid~\cite{feng2022gifdroid} utilized image processing techniques to extract key information from screen recordings and then automate bug reproduction.
Wang et al.~\cite{wang2025empirical} made a systematic study on the images within bug reports and found that using images in bug reports can enhance the reproduction task.
We leave the reproduction of image-based bug reports as future work.

\oursystem inputs the whole bug reports into an advanced LLM to extract more comprehensive information, without the use of S2R entities and specific bug type domains.
Meanwhile, it feeds both textual reports and UI screenshots into the LLM to facilitate decision-making by leveraging their complementarity.
Moreover, by integrating MCTS with LLMs, our approach achieves a more strategic and guided searching of the state space of apps. 
This synergy leads to noticeably better outcomes, as confirmed by our experiments.

\subsection{LLM-based Bug Report Reproduction}
With the success of LLMs in NLP tasks, researchers have begun to explore their potential in automating bug report reproduction.
AdbGPT~\cite{feng2024prompting} first leveraged prompt engineering to reproduce bugs without requiring any training or hard-coded rules.
However, this tool faces two limitations: (i) difficulty in accurately extracting S2R entities from bug reports with complex and diverse semantics~\cite{wang2024feedback}, and (ii) limited long-term planning due to prompt-only reliance.
ReBL~\cite{wang2024feedback} bypassed the use of S2R entities and incorporated a feedback mechanism into LLM prompts, yet still struggled in complex UI exploration.
ReActDroid~\cite{huang2025one} reproduced app bugs from the crash overview of one-sentence reports.
It leveraged the LLM with ReAct prompting to iteratively interact with the GUI widget that may lead to the crash.
As with ReBL, prompt-only approaches may not always be sufficient to ensure effective and reliable bug reproduction.
AndroB2O~\cite{johnson2025androb2o} is an LLM-based approach that generates test oracles for non-crashing GUI failures by reasoning over multimodal bug reports combining text and screenshots.
Unlike our method, which targets crash-inducing bugs and leverages screenshots at every interaction step, AndroB2O focused on non-crashing GUI failures and used only the screenshot related to the failures.
AEGIS~\cite{DBLP:conf/sigsoft/WangGM0HLG25} and \textsc{LIBRO}~\cite{kang2023large} used LLMs for code-level bug reproduction, rather than addressing the challenges arising from UI interactions.

To better understand UI states, \oursystem takes multimodal inputs, \eg, the complete bug report and the UI screenshot captured during reproduction, rather than relying on explicit S2R entity extraction.
Inspired by the advanced techniques above, it integrates an LLM-based multi-agent framework with MCTS for effective state space searching: MCTS guides decision-making, while two LLM agents respectively identify top-\textit{k} promising actions and estimate their success likelihood after executing each selected action.


\subsection{Bug Report Study}
Several research efforts have focused on studying and analyzing Android bug reports.
\textsc{Yakusu}~\cite{fazzini2018automatically} combined program analysis and NLP to generate executable test cases from bug reports.
MaCa~\cite{liu2020automated} trained a machine learning-based classifier to identify and classify action words in the bug reports of apps.
Johnson et al.~\cite{johnson2022empirical} conducted an empirical study to examine the challenges of bug reproduction and the quality of the reported information.
\textsc{Burt}~\cite{song2022toward} developed a task-oriented chatbot with instant feedback and graphical suggestions to improve the quality of bug reports.
EBug~\cite{fazzini2022enhancing} helped users write more accurate bug reports by connecting reproduction steps with relevant information obtained from static and dynamic analyses of the app.
TAB~\cite{liu2025tab} was designed to generate accurate and meaningful titles for bug reports automatically.
\textsc{BugSpot}~\cite{zhang2025automated} automatically recognized the buggy behavior of the described bug during the automated reproduction of the bug report.
It could be combined with our work to enhance the overall effectiveness in the future.

Although these techniques offer valuable insights into understanding bug reports, their primary objectives are not to directly reproduce the reported bugs.
Nevertheless, the techniques on analyzing and improving bug reports can benefit \oursystem, for example by enhancing its understanding of the semantic of bug reports.

\section{Conclusion}
We introduce \oursystem, a novel technique that combines LLM reasoning with MCTS-based planning to reproduce Android crashes from incomplete bug reports.
Unlike prior LLM-based approaches that rely solely on prompt engineering, \oursystem treats bug reproduction as a goal-directed search task and leverages two LLM-guided agents for semantic reasoning and adaptive decision-making within MCTS.
Extensive experiments on 93 real-world Android bug reports show that \oursystem achieves a reproduction success rate of 64.52\%, significantly outperforming four state-of-the-art baselines.
Each successful reproduction takes an average of 241.4 seconds, which is acceptable in practical scenarios.

\bibliographystyle{ACM-Reference-Format}
\bibliography{sample-acmsmall-conf.bib}

\end{document}